\documentclass{article}

\usepackage{graphicx} 
\graphicspath{ {images/} }

\usepackage[rightcaption]{sidecap}

\usepackage{wrapfig}

\begin{document}

\title{Perpetual generation of two-photon quantum beats}

\author{C. Dedes\thanks{{c\_dedes@yahoo.com}}\\
Bradford College  \\
Great Horton Road, Bradford \\
West Yorkshire, BD7 1AY \\
United Kingdom \\}

\maketitle

\begin{abstract} We consider a recursive device that is based on a Mach-Zehnder interferometer and linear optical elements which allow self-feedback through multiple internal reflection of radiation between two parallel arrays of opposite faced mirrors. By a carefully chosen experimental arrangement and for certain input states it is possible to observe at the open ends of the device time generated coherent superpositions \textit{in perpetuum}.
\end{abstract}

It is generally accepted that a number of genuine quantum optical effects are related to higher order correlation functions \cite{Hong} and a rigorous quantum treatment of one of them, the celebrated Hanbury-Brown and Twiss effect (HBT) \cite{Fano}, reveals the more subtle properties of the radiation field. Related correlation effects have also been studied in nuclear  and condensed matter physics \cite{Baym}. A qualitative explanation of the HBT effect relies on the indistinguishability of the possible ways that two photons may be detected by two separated detectors \cite{Fano}. The interference of the these two amplitudes gives rise to the intensity correlation fringes.  Historically radiation escaping cavity.cavity leaking \cite{Kuhn} .Conceptually challenging two-particle interference effects have been discussed even from the early days of wave mechanics \cite{Darwin}. Diffraction in the time domain has been illustrated by a set-up put forward by Moshinsky \cite{Moshinsky,Singlephoton}. More recently time related interference effects \cite{Czachor}. Ghost imaging experiments \cite{Sergienko} have also indicated a novel way of creating spatial correlations and interference fringes.

The main idea in this paper is the interference of two alternative histories amplitudes (see Fig. 1 for an outline of the set up). Two photon wavepackets are inserted from two different inputs of a three-port device as shown. One amplitude is related to the process of one of the photons arriving at one of the detectors after having travelled one time inside the interferometric device while the other photon is reflected one more time and travels twice in the interferemeter before being detected at the output (b). The other amplitude refers to a similar process but this time the two photons are absorbed by detectors 1 and 2 respectively (we assume we put the detectors in place just before the arrival of the photons). Since the photons are identical these processes cannot be distinguished so the corresponding probability amplitudes should be added together. We see then a type of interference between two identical processes and not between two overlapping photon wavepackets (the photons are spatially separated and follow different paths). This kind of interference between two (or more) photons is new and certainly does not support Dirac's famous remark \cite{Hong}. 
 
A significant notion of this thought experiment is that the particular conditions of the configuration decided by the observer ahead of time is of crucial importance for the observation of the effect. If instead, the experimentalist had decided to place both the detectors at different points, the type of intensity interference explained earlier would be absent since both photons would definitely travel twice inside the interferometer before being detected (so there would be no ambiguity that allows interference of amplitudes). The time we choose for the measurements is also very important as it is apparent from the previous discussion. The influence of the particular experimental conditions for the available type of possible predictions has been emphatically argued by Bohr\cite{Murdoch}. 

Here we discuss a thought experiment that illustrates the effect of time in higher order interference. We consider a linear optical set up \cite{KLM}, the device of Fig. (1). A phase shifter $\phi/2$ induces a path length difference between the interfering paths comparable to the coherence lengths of the wavepackets. Since a continuum mode description of photon radiation is more realistic and conceptually richer, it enables us to study in greater detail the spatial-temporal profile of photons (see for example the celebrated Hong-Ou Mandel dip effect). We proceed then by considering a two-photon wavepacket input and by adopting the space-time domain description in \cite{Legero} (see also \cite{Blow}). After the first reflection the electric field raising operators at the output are expressed as a linear transform \cite{Reck} of the incoming mode functions

\begin{equation}
    \hat{A}_{1}(t)=\frac{1}{\sqrt{2}}\left[-\zeta_{1}(t)\hat{a}_{1}+\zeta_{2}(t)\hat{a}_{2}\right],
\end{equation}

\begin{equation}
    \hat{A}_{2}(t)=\frac{1}{\sqrt{2}}[\zeta_{1}(t)\hat{a}_{1}+\zeta_{2}(t)\hat{a}_{2}],
\end{equation}

\noindent
where

\begin{equation}
    \zeta_{1,2}(t)=e^{-i\theta_{1,2}\left(t\right)}\epsilon_{1,2}\left(t\right),
\end{equation}

\noindent
and $\epsilon_{i}$ Gaussian normalized envelope functions. After the second reflection the photons arrive at detectors 1 and 2 simultaneously and the coincidence probability is given by

\begin{eqnarray}
  P_{2'3}=\frac{1}{2^{11}}\bigg|\left[\zeta_{1}(t_{0}+\tau)-\zeta_{1}(t_{0}+\tau+\delta \tau)\right ]\left[ 
-\zeta_{2}(t_{0}+\tau)+\zeta_{2}(t_{0}+\tau+2\delta \tau)\right] \nonumber \\
+ \left[\zeta_{2}(t_{0})-\zeta_{2}(t_{0}+\delta \tau)\right ]\left[-\zeta_{1}(t_{0}+\tau)+\zeta_{1}(t_{0}+\tau+2\delta \tau)\right]  \bigg|^{2},
\end{eqnarray}

\noindent
where $\delta \tau$ the time differences induced by a phase shift $\phi$. When $\delta\tau=0$ the coincidence rate is zero even if $\tau \neq 0$ so the photons enter the device at different times. The physical interpretation of (4) is appparent from its form which clearly illustrates that when one photon wave-packet travels twice through the device the second one travels only once since there are two instinstiguishable ways for this to happen fourth order interference is generated. Obviously, we could vary the phase $\phi$ and consequently the time $\delta \tau$. Based on (4) we can examine the conditions for possible violation of the Cauchy-Schwarz or Bell-type inequalities, for population trapping etc. We may also need to to calculate higher correlation functions in a multiport device setting \cite{Zhang} (allowing at least two spatial modes to interfere). It is crucially important to note the presence of cross intereference terms in the above relation despite the fact that, loosely speaking, one photon is reflected just once and the other twice before both being absorbed by the spatially separated detectors. To the contrary it is exactly the positioning of the detectors and the time we perform the measurements along with the indistinguishability of the two photons that give rise to these intensity interference fringes, since the probability amplitudes for the two indistinguishable processes add up. It should be added furthermore that since we are dealing with a nonclassical state input the output must be entangled \cite{Kim}. For a coherent (or thermal) state on the other hand the output is a product state and there are no nonclassical correlations between the modes. We may also say that the Hilbert space of the device expands as a part of the radiation is reflected and fed back into the device.

Possible realizations of the proposed thought experiment may be possible using parametric down converters \cite{Hong} or correlated spontaneous two photon emission from cascade atoms \cite{Fearn}. It might be beneficial from a calculational point of view to examine the same type of problems using characteristic functions and quasi-probability distribution densities in phase space \cite{Blow},\cite{Narducci}. Here, the Janszky representation \cite{Janszky} as illustrated recently by Barnett \cite{Barnett} may be of considerable utility since it simplifies the calculations involved as it relies on coherent states. The Heisenberg picture description of the presented device may be not appropriate since a part of the output enters again in the device (see also the relevant discussion in "\cite{Franson2}). Increasingly more complicated superpositions will be produced by more general multimode impinged input states so a photon aggregate fed in the device will create increasingly complicated inter-mode higher order correlations between photons at different radiation fronts. It must be added that even negative measurements  will have an effect on the properties of the radiation field. It may be noted that it is not easy to invert the unitary transformation made by the device even though all the elements are non-absorving. A possible inversion of the photon momenta by using simple reflection at the output will not suffice as it does in most common interferometers since as time evolves new superpositions are generated (even though the Hamiltonian of the device is time-independent).

In conclusion, we have described an interferometric device with perpetual internal reflection and some leaky interfering modes that allow multipath coherence and under certain conditions give rise to a new type of intensity correlations. The core physical idea behind our scheme is that by a careful experimental arrangement we create the the right conditions for the interference of probability amplitudes which describe indistinguishable histories. In that way, certain superposition of states are created with the help of this device as time passes, something that is not encountered in simple or multiport Mach-Zehnder type of devices \cite{Reck} the state of the radiation field becomes progressively extremely complicated as the photons follow all the available indistinguishable histories. The above setting may be extended also for other boson fields or even for fermion interferometry experiments \cite{Loudonferm}.

  \begin{figure}
    \includegraphics[scale=0.4]{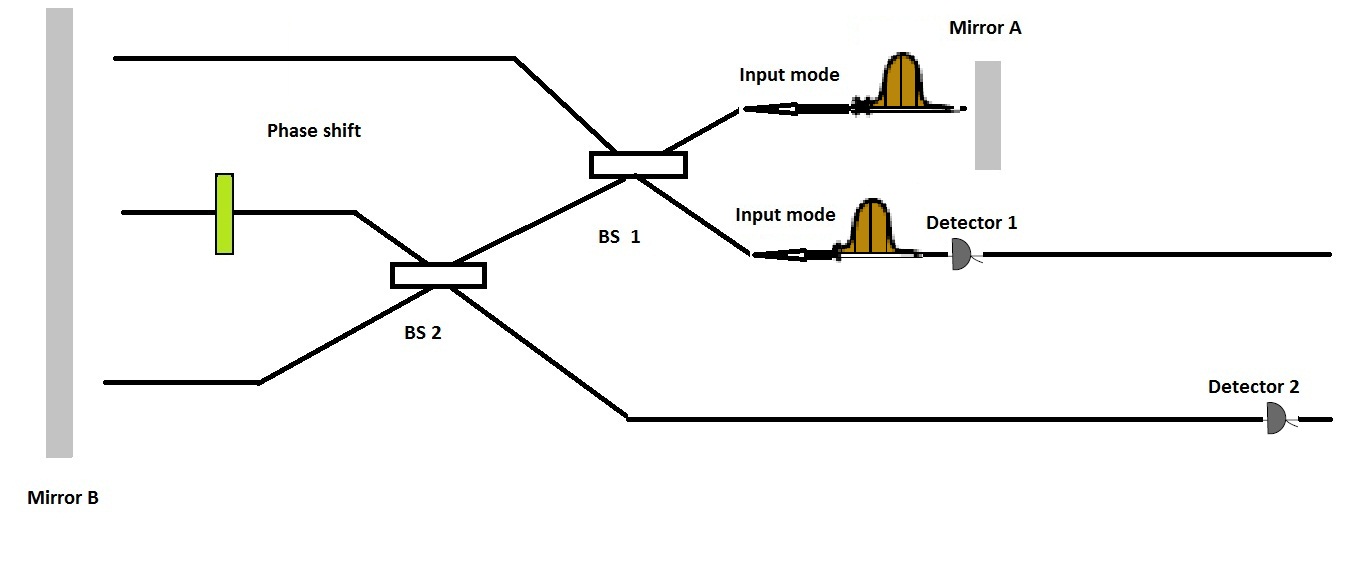}
     Schematic representation of the proposed thought experiment. Two photons enter a multiport beam splitter configuration through different modes with a time delay $\tau$ between their wavepackets and detected at the same time at the output detectors 1 and 2. The constant phase shift induces a time delay $\delta\tau$. One of the photons is reflected once by the left side array of mirrors and arrives at detector $1$ while the other is reflected by the right hand mirror, re-enters the device, is reflected again from the left side array and finally is absorbed by detector $2$. Another possible process follows by interchanging the paths of the two photons. Since these two processes cannot be distinguished the corresponding amplitudes should be added and this is the physical reason for the existence of two-photon interference.
\end{figure}


\begin{thebibliography}{99}

\bibitem{Hong}
 C. K. Hong, Z. Y. Ou, and L. Mandel, \emph{Phys. Rev. Lett.},\textbf{59}, 2044 (1987);
 J. D. Franson. \emph{Phys. Rev. Lett.}, \textbf{62}, 2205 1989;
 X. Y. Zou, L. J. Wang, and L. Mandel. \emph{Phys. Rev. Lett.}, \textbf{67},318 (1991);
 D.M. Greenberger, M.A. Horne, and A. Zeilinger. \emph{Phys. Today}, \textbf{46}, 22 (1993);
 Z. Y. J. Ou. \emph{Multi-Photon Quantum Interference}, Springer, (2007).
 
\bibitem{Fano}
  U. Fano, \emph{Am. J. Phys.}, \textbf{29}, 539 (1961);
  M.O. Scully and M.S. Zubairy. \emph{Quantum optics}. Cambridge University Press, (1997);
  R.J. Glauber. \emph{Quantum Theory of Optical Coherence}. Wiley, (2007).
  
  \bibitem{Baym} 
 G. Baym. \emph{Acta Phys. Polon. B}, \textbf{29}, 1839, (1998);
 P. Samuelsson,  E.V. Sukhorukov,  and M. Buttiker.   \emph{Phys. Rev. Lett.}, \textbf{92}, 026805 (2004);
 I. Neder, N. Ofek, Y. Chung, M. Heiblum, D. Mahalu, and V. Umansky. \emph{Nature}, \textbf{448}, 333 (2007).
  
  \bibitem{Kuhn}  T. S. Kuhn, \emph{Black-Body Theory and the Quantum Discontinuity, 1894-1912}, 1987 (University of Chicago Press, second edition); Max Planck, \emph{The Theory of Heat Radiation},  F. Blakiston Son  Co., 1914 
  
 \bibitem {Darwin}  C. G. Darwin, Proc. R. Soc. Lond. A, \textbf{124}, 375 (1929)
 
  
\bibitem{Moshinsky} M. Moshinsky, \emph{Phys. Rev.} \textbf{88}, 625 (1952)

\bibitem{Singlephoton}  B. Lounis, M. Orrit, \emph{Rep. Prog. Phys.} \textbf{68} 1129-1179 (2005)

\bibitem{Czachor} M. Czachor, \emph{Phys. Lett. A} \textbf{383}, 2704-2712 (2019).

\bibitem{Sergienko} Simon, D.S., Jaeger, G., Sergienko, A.V., Quantum Metrology, Imaging, and communication, Springer (2016)

 
  \bibitem{Murdoch} D. R. Murdoch, Bohr's philosophy of quantum mechanics, Cambridge University Press (1989)

\bibitem{KLM}
  E. Knill, R. Laflamme, and G. J. Milburn, \emph{Nature}, \textbf{409}, 46 (2001);
  P. Kok, W. J. Munro, K. Nemoto, T. C. Ralph, and Dowling J. P. \emph{Rev. Mod. Phys.}, \textbf{79}, 135 (2007);
  P. Kok and B. W. Lovett. \emph{Introduction to Optical Quantum Information Processing}, Cambridge University Press (2010).
  
   \bibitem{Legero}
  T. Legero, T. Wilk, A. Kuhn, and G. Rempe. \emph{Applied Physics B: Lasers and Optics}, \textbf{77}, 797, (2003).
  
  \bibitem{Blow}
  K. J. Blow, Rodney Loudon, Simon J. D. Phoenix, and T. J. Shepherd, \emph{Phys. Rev. A}, \textbf{42},4102 (1990);
  S.M. Barnett and P.M. Radmore. \emph{Methods in theoretical quantum optics}, Oxford University Press (1997) ;
  R. Loudon. \emph{The quantum theory of light}, Oxford University Press (2003);
  A. M. Brańczyk, (2017) arXiv:1711.00080 [quant-ph].
  
\bibitem{Reck}
 M. Reck, A. Zeilinger, H.J. Bernstein, and P. Bertani, \emph{Phys. Rev. Lett.}, \textbf{73}, 58 (1994);
 A. Vourdas and J. A. Dunningham, \emph{Phys. Rev. A}, \textbf{71}, 013809 (2005).
 
\bibitem{Zhang}
  S. Zhang, C. Lei, A. Vourdas, and J. A. Dunningham. \emph{J. Phys. B: At.Mol. Opt. Phys.}, \textbf{39}, 1625 (2006).

\bibitem{Kim}
  M. S. Kim, W. Son, V. Buzek, and P. L. Knight, \emph{Phys. Rev. A}, \textbf{65}, 032323 (2002).
  
\bibitem{Fearn}
  Fearn H. and R. Loudon, \emph{J. Opt. Soc. Am. B}, \textbf{6}, 917 (1989);
  U. W. Rathe and M. O. Scully. \emph{Letters in Mathematical Physics}, \textbf{34}, 297 (1995).
  
\bibitem{Narducci}
  Lee E. Estes, Thomas H. Keil, and Lorenzo M. Narducci, \emph{Phys. Rev.}, \textbf{175}, 286 (1968);
  Z. Y. Ou, C. K. Hong, and L. L. Mandel. \emph{Opt. Commun.}, \textbf{63}, 118  (1987).


\bibitem{Janszky}  J. Janszky and A.V. Vinogradov, \emph{Phys. Rev. Lett.} \textbf{64}, 2771 (1990)

\bibitem{Barnett} S. M. Barnett, \emph{Journal of Russian Laser Research}, \textbf{39}, Issue 4, pp 340–348 (2018)


\bibitem{Franson2} J.D. Franson and R.A. Brewster: Limitations on the use of the Heisenberg picture (2018) arXiv: 1811.06517 [quant-ph]


\bibitem{Loudonferm}
  R. Loudon. \emph{Phys. Rev. A}, \textbf{58}, 4904 (1998).
  
\end{thebibliography}
\end{document}